\documentstyle[aps,multicol,eqsecnum,epsf]{revtex}
\begin{document}
\draft
\title{Effects of Defects on Friction for a Xe Film Sliding on Ag(111)} 
\author{M. S. Tomassone and J. B. Sokoloff}
\address{Center for Interdisciplinary Research on Complex Systems and 
Department Physics, Northeastern University, Boston, MA 02215}
\date{\today }
\maketitle

\begin{abstract}

The effects of a step defect and a random array of point defects (such as 
vacancies or substitutional impurities) on the force of friction acting on 
a xenon 
monolayer film as it slides on a silver (111) substrate  are studied by 
molecular dynamic simulations and compared with the results of lowest order 
perturbation theory in the substrate corrugation potential. For the case of a 
step, the magnitude and velocity dependence of the friction force are strongly 
dependent on the direction  of sliding respect to the step and the 
corrugation strength. When the applied force F is perpendicular to the
step, the film is pinned for 
F less than a critical force $F_{c}$. 
Motion of the film along the step, however, is not pinned. Fluctuations in 
the sliding velocity in time provide evidence of both stick-slip motion 
and thermally activated creep. Simulations done with a substrate containing a 
5 percent concentration of random point defects for various directions of 
the applied force 
show that the film is pinned for the force below a critical value. 
The critical force, however, is still much lower than the effective inertial 
force exerted on the film by the oscillations of the substrate in 
experiments done with a quartz crystal microbalance (QCM).  
Lowest order perturbation theory in the 
substrate potential is shown to give results consistent with the simulations, 
and it is used to give a physical picture of what could be expected for real 
surfaces which contain many defects. 

\end{abstract}

\section{Introduction}
\begin{multicols}{2}
\narrowtext

In previous simulations of a xenon film sliding on a silver substrate, 
using a periodic (i.e., defect-free) 
substrate, a viscous force of friction 
(i.e. one proportional to the sliding velocity) was found
\cite{robbins,tomassone}, in agreement with the 
experimental results of Krim  et. al. \cite{krim1,daly}. In contrast, 
perturbation theory calculations give a velocity independent contribution to 
the friction 
(i.e., "dry friction") when there are point defects (i.e., a point defect 
denotes a defect which is centered around a point in the lattice), such as 
vacancies or substitutional impurities \cite {sokoloff,defect1,soko-toma} 
in the substrate.
Since, real surfaces, even very smooth ones, always 
contain defects, the viscous friction found in these 
experiments\cite{krim1,daly} 
is a surprising result. Perturbation theory results for straight line defects, 
i.e., defects that extend along a line 
such as steps or facet boundaries\cite {soko-toma}, however, 
are consistent with 
viscous friction. In this article, 
we perform molecular 
dynamics simulations for a film of Xe atoms sliding on a 
Ag(111) substrate containing a step and also for a substrate with a random 
array of point defects. The point defects are found to pin the film 
for an applied force below a critical value, whereas a step generally does 
not pin the film. The pinning force due to a 5 percent concentration of 
point defects comparable in strength to the corrugation potential, however, is 
still much lower than the effective inertial force per film atom in the 
experiment done by krim.\cite{krim1,daly}  
thIs force can be expressed as $m\omega^2 A$, 
where m is the atomic mass and 
$\omega$ and A are the frequency and amplitude, respectively, of the 
quartz crystal microbalance used in the experiment. Lowest order 
perturbation theory 
in the substrate corrugation potential for the system treated 
in the simulations is found to be consistent with the simulations. 
The perturbation theory results are then used to try to give a physical 
picture of what one would expect for real substrate surfaces, which contain 
many defects. 

\section{Simulations}

\subsection{The model used in the simulations}

The model Hamiltonian used in Ref. \cite{tomassone} for $N$ film atoms of mass $m$ 
at positions ${\bf r}_k$ $%
(k=1,..,N)$ is given by 
\begin{equation}
\label{hamilton}H \equiv \sum_{k=1}^{N} \frac{{\bf p}_k^2}{2m}
+U({\bf r}_1,\cdots ,{\bf r}_N), 
\end{equation}
where ${\bf p}_k$ is the momentum of the atom $k$, and the total potential $%
U({\bf r}_1,\cdots ,{\bf r}_N)$ is given by 
\begin{equation}
U({\bf r}_1,\cdots ,{\bf r}_N)\equiv \sum_{k=1}^{N}U_s({\bf r}%
_k)+\sum_{j<k=1}^{ N}V(|{\bf r}_j-{\bf r}_k|). 
\end{equation}
Here, $U_s({\bf r}_k)$ is a single particle potential describing the
interaction between the $k$-th film atom and the substrate, and $V(|{\bf r}%
_j-{\bf r}_k|)$ is the pair potential interaction between the $j$-th and $k$%
-th atoms in the film.

The interaction between two Xe atoms is given by a Lennard-Jones potential
\begin{equation}
V(r)=4\varepsilon \left[ \left( \frac \sigma r\right) ^{12}-\left( \frac 
\sigma r\right) ^6\right] , 
\end{equation}
where $\varepsilon =19.83\,meV$, and $\sigma =4.055\,\AA.$ The
interaction between a Xe atom and the substrate can be  described
by  a  substrate potential without internal degrees of freedom 
given by \cite{steele}
 
\begin{equation}
\label{us}U_s({\bf r}_{\Vert },z)=U_0(z)+U_1(z)\sum_{\{{\bf G}\}}\cos ({\bf %
G\cdot r}_{\Vert }), 
\end{equation}
where ${\bf r}_{\Vert }=(x,y)$ are the coordinates of the Xe atom parallel
to the substrate, and $\{{\bf G}\}$ is the set of the six shortest
reciprocal lattice vectors of the substrate. The first term in Eq. (\ref{us}%
) describes the mean interaction of the atoms with the substrate, and the
second term describes the periodic corrugation potential.

Expressions for $U_0(z)$ and $U_1(z)$ were derived by Steele \cite{steele}
assuming that the substrate potential $U_s({\bf r})$ is  a sum of
Lennard-Jones  potentials between one film atom and
all of the atoms in the substrate.  However, a potential like $U_s({\bf r})$, 
which is a sum of Lennard-Jones potentials is not a correct
description of the interaction of a metallic surface with a noble gas atom. The
corrugation potential is reduced (from the value found by summing
Lennard-Jones potentials) due to electronic screening. For this 
reason we employ a weaker corrugation potential, 
as did Cieplak {\it et al.} in Ref. \cite{robbins}.
The corrugation potential we use is

\begin{equation}
\label{pote2}U_1(z^{*})=\alpha e^{-g_1z^{*}}\sqrt{\frac \pi {2g_1z^{*}}}\left[ 
\frac{A^{*6}}{30}\left( \frac{g_1}{2z^{*}}\right) ^5-2\left( \frac{g_1}{2z^{*}}%
\right) ^2\right] , 
\end{equation}
where $\alpha =4\pi \varepsilon _{Xe/Ag}A^{*6}/\sqrt{3}$, 
$z^{*}=z/a$, $a=2.892$ $\AA$ 
is the lattice constant of the substrate, $A^{*}=\sigma _{Xe/Ag}/a$, 
$g_1=4 \pi/ \sqrt{3}$.
We calculate the Lennard-Jones parameters $\sigma _{Xe/Ag} $ 
and $\varepsilon _{Xe/Ag}$ by  fitting 
(i) the position of the minimum of $U_{0}(z)$ 
to the distance between a Xe atom in the first layer and the ion  
cores of the substrate ($z_{0}$ ), and 
(ii) the attractive
well depth to the binding energy of one $Xe$ atom to the $Ag$ substrate
$(U_{0}(z_{0})=-211$ $meV$,  from \cite{cole}).
We find  $\sigma _{Xe/Ag}=4.463\,\AA $ and 
$\varepsilon _{Xe/Ag}=13.88\,\,meV$.

The corrugation potential $U_1(z)$ in Eq. (\ref{pote2}) falls off exponentially at large z. The above 
parameters give $U_{1}(z_{0}^{*})\sum_{\{{\bf G}\}}\cos ({\bf %
G\cdot r}_{\Vert })=2.025$ $meV$ (for the maximum value of this sum)
in contrast to the corresponding 
value of $10.13$ $meV$ for Steele's potential at $z_{0}$ \cite{steele}. 
Our corrugation gives good agreement with the experimental value of the 
slip time.

Our simulations are carried out at an equilibrium temperature of 
$T=77.4\,^oK $, and the particles move in a three dimensional box of
size  $ 20\,a \times $ $10\,a\sqrt{3}\times 10\,\sigma$. 
The time scale for vibrations of the adsorbed film
atoms is $t_0=\sqrt{(m\sigma ^2/\varepsilon )}$ $=3.345$ $ps$ , with 
$m=2.16$ $10^{-22}$ $g$. Periodic
boundary conditions in the $x$ and $y$ directions are employed along with a
hard wall boundary condition in the $z$ direction at the top of the box.

We change the coverage by changing the number of Xe atoms $N$.
We use $60\leq N\leq 370$.
All atoms are initially in the gas phase. The atoms condense in $250\,t_0$\ or
less, forming a triangular lattice incommensurate with the substrate
fcc(111) surface. 

\begin{figure}
\centerline{
\vbox{ \hbox{\epsfxsize=4.5cm \epsfbox{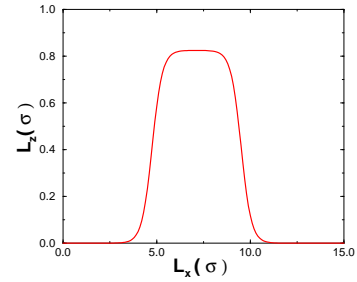}}
       \vspace*{1.0cm}
        } 
}
\caption{A plot of $g(x)= 0.58\sigma [f(x_1-x)-f(x-x_2)]$, where $f(x)$ 
is the Fermi function.}
\label{fermi}
\end{figure}

For the simulations done with a step present, the potential in equation 
(\ref{us}) has the z-coordinate replaced by 
$z-g(x)$. The use of this function guarantees that we have
the same corrugation as we used for a substrate free of defects 
in Ref. \cite{tomassone}. 
 Here $g(x)= 0.58\sigma [f(x_1-x)-f(x-x_2)]$ where  $f(x)$ 
is the Fermi function, $1/[e^{-x /\omega}+1]$. (See Fig. \ref{fermi}.)
We choose $\omega$ to be equal to 
$1.1$ $\sigma$, AS THE  width for the step edge, where $\sigma$ is the 
distance parameter for the Lennard-Jones 
potential between a film and substrate atom. 

\begin{figure}
\centerline{
\vbox{ \hbox{\epsfxsize=4.5cm \epsfbox{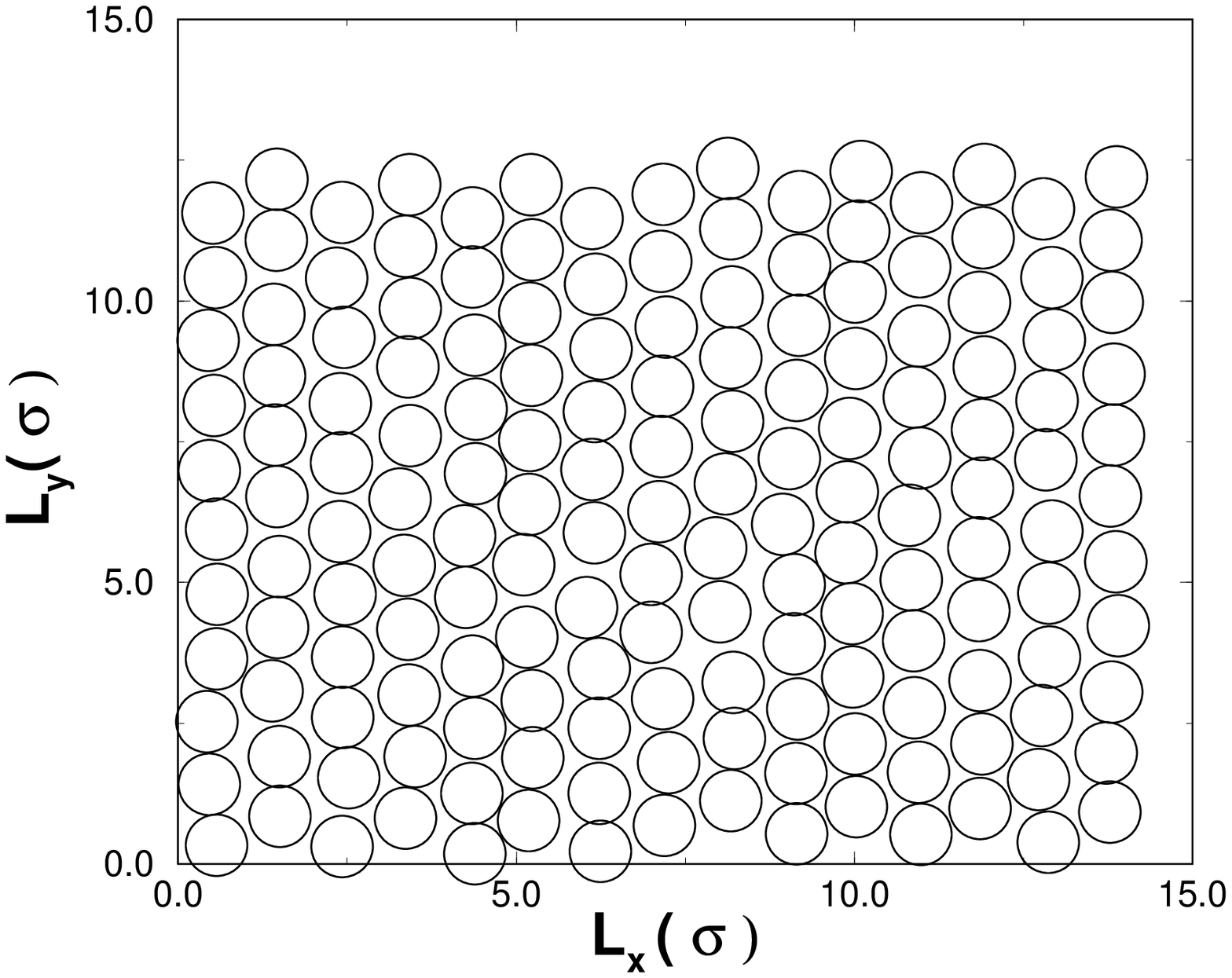}
                \hspace*{0.5cm} \epsfxsize= 4.5cm  \epsfbox{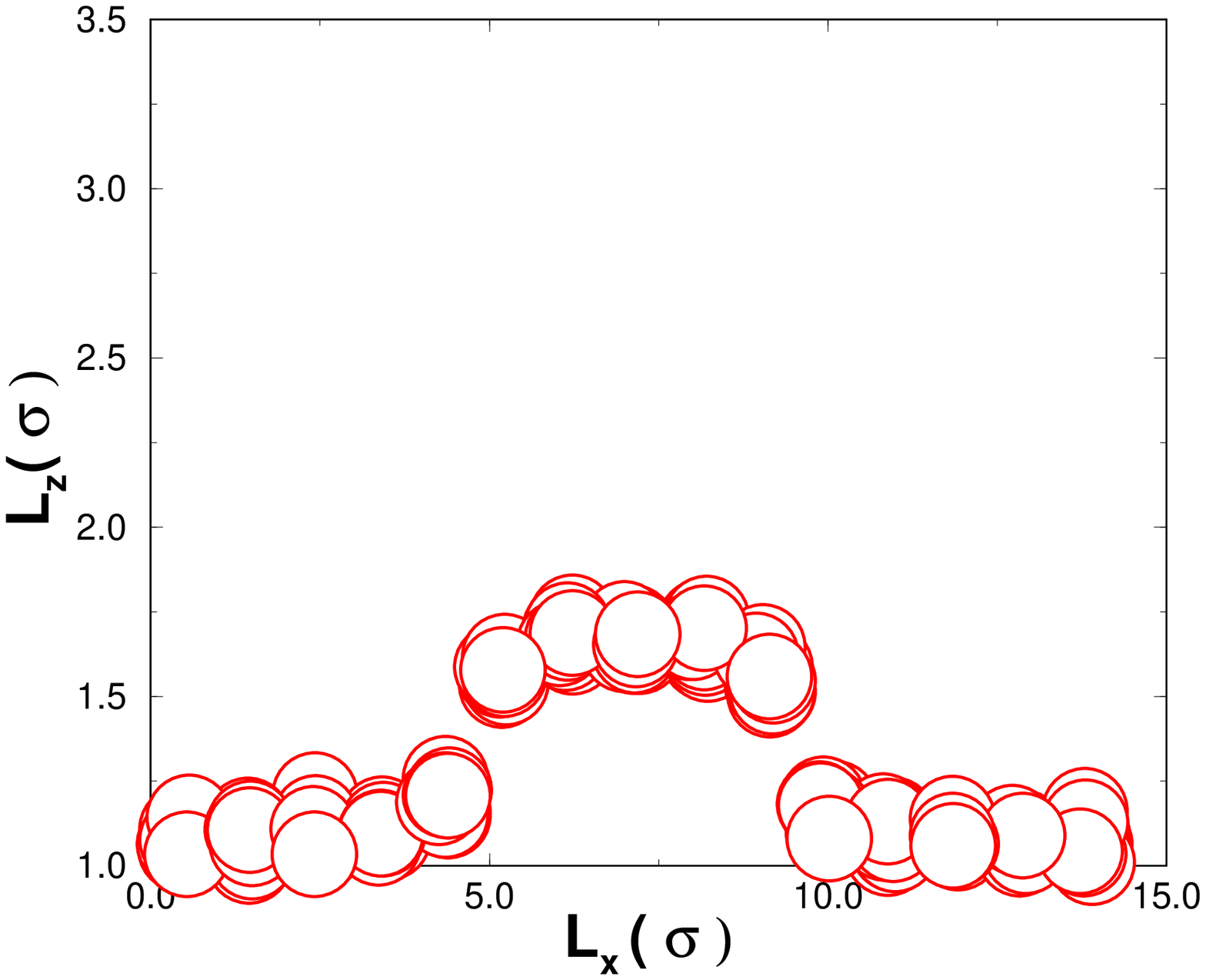} }
       \vspace*{1.0cm}
        } 
}
\caption{a. Upper view of the position of the particles after 200,000 
iterations of the program. The step 
is located along the $y$ axis between $5$ and $10$ $\sigma$. 
b. Side view of the same sets of positions.}
\label{position}
\end{figure}

\noindent
We take $x_1$ and $x_2$, the locations of the beginning 
and end of the step, equal to $ L_{x}/3$ and $2L_{x}/3$, where $L_{x}$ 
is the length of the 
box (along x) in which the simulations are done. 
We thus assume a straight step in the substrate, (see Fig. \ref{position} 
(a and b)),
which  runs along the y-axis.
The height of the step is about 0.8 of an atomic distance. The use of a 
function $g(x)$, which varies smoothly 
with $x$, is a reasonable choice because the nonzero radius of a surface 
atom makes the potential that acts on the atom vary smoothly 
as the atom moves over the surface.

Most of our simulations were done with a coverage corresponding to the 
uncompressed monolayer 
(163 particles).  Periodic
boundary conditions in the $x$ and $y$ directions are employed along 
with a
hard wall boundary condition in the $z$ direction at the top of the box.
Because 
of our use of periodic boundary conditions, we are technically 
simulating a 
periodic array of defects. We feel, however, that the cell length used in our 
simulations of about 13 xenon atom spacings is  sufficiently
long for the coherence from one cell to the next to be unimportant for most 
values of the sliding velocity. This 
is justified using perturbation theory\cite {sokoloff,defect1,soko-toma} 
for reasonable values of the phonon damping constant in appendix A.

In the present simulations, we use the same method as Cieplak, et. al. 
\cite{robbins},  in 
which a constant external force $F$ is applied to each atom in the plane 
of sliding, and the 
resulting steady-state velocity 
is calculated (see Fig. \ref{velocity}). Throughout the simulation a 
thermostat that rescales the three velocity components is used to maintain 
constant temperature.  
The rescaling is done in the center of mass reference frame so as not 
to change the center of mass velocity 
and thus introduce an 
unphysical 
force of friction 
due to the thermostat. It renormalizes the 
atomic velocities every time step, so that the total kinetic energy 
per atom in the center of mass frame 
is maintained at $(3/2)k_B T$, where $k_B$ is Boltzmann's constant and $T$ 
is the desired absolute temperature of the system ($T=77.4\,^oK $ in 
our case). This method allows us  to determine conveniently 
the velocity dependence of the friction force. 
The velocity of the film is affected by the inhomogeneities of the medium,
in this case the corrugation and the defects. In the absence of defects, the 
film can move for an arbitrarily weak applied force, but when there 
is a step 
present, the film can be pinned for $F$ less than a critical value $F_c$, 
if the force is applied perpendicular to the step. 
In other words, there are the following two phases, 
a {\bf pinned phase} when the external force $F$ is $F<F_{c}$, and a 
{\bf moving phase}. 
The {\bf depinning transition} 
takes place at a critical
threshold force $F_{c}$. 
In the vicinity of the depinning transition 
the average velocity has the form $V \sim (F-F_{c})^{\beta}$ where 
$\beta$ is the velocity  exponent.  

\begin{figure}
\centerline{
\vbox{ \hbox{\epsfxsize=4.5cm \epsfbox{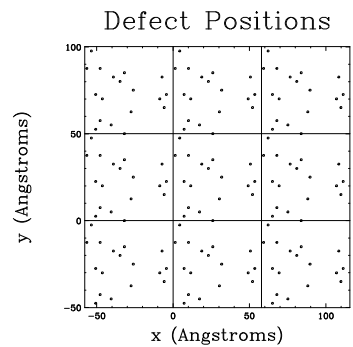}}
       \vspace*{1.0cm}
        }
}
\caption{The defect positions are shown. The central box is the cell in which 
the simulations were performed, and the neighboring cells represent the 
reflected defect positions.}
\label{imp}
\end{figure}

The potential for a substrate 
containing point 
defects is of the form:
\begin{equation}
\label{imppot}
V_i (\vec r-\vec r_d)=V_0 e^{-|\vec r-\vec r_d|^2/\ell^2}
\end{equation}
where $V_0$ and $\ell$ are the strength and range of the potential respectively
and $\vec r_d$ is the location of the defect. The defect positions 
are chosen 
to be potential minima of the corrugation potential chosen at random with 
probability $c$, where $c$ is the defect concentration. 
(The simulations done for 
point defects were done with $U_1$ having the opposite sign than for the 
$U_1$ used in the step simulations, because this sign of $U_1$ gives the 
correct corrugation potential minima. For the step simulations it did not 
really matter what sign we chose, but for the point defect problem it is 
more important to place the point defects at the correct positions. 
For example if we want $V_i$ to represent the potential due to a vacancy we 
can choose $V_0$ positive and equal to the depth of the corrugation 
potential minimum, so that that minimum, presumed to be due to a substrate 
ion located at that position, is canceled out by $V_0$.) 
Because of the use of periodic boundary conditions, it was necessary to 
include the eight nearest neighbor cells to the cell containing the 
film in which the simulations were done. Each of these neighboring cells 
contained 
the defect potentials reflected into these cells. The impurity positions and 
the reflected impurity positions are shown in Fig. \ref{imp}. Only interactions 
between a film atom and a defect (or an image of a defect) which is less than 
$4\sigma$ away from the atom are included in the calculation of the force 
or potential acting on the atom.

\subsection{Results of the Simulations for the Step}

In most of our simulations we have used a corrugation strength 
(defined as 9 times the value of $fU_1 (z)$ evaluated at the value of 
z at which $U_0(z)$ is minimum)  of 2.025 meV ( f=1), the value that was 
used in Ref. \cite {tomassone}. In order to examine the effect of using 
a larger corrugation, we have, in addition, done some runs at a corrugation 
of 3.56 meV (f=1.69).

We apply an external force to all the particles in the monolayer 
at the following angles with respect to the $x$ axis (which is perpendicular 
to the step): 
$0 ^{o}$ (force perpendicular to the step), 
$16 ^{o}$, $37 ^{o}$, $58 ^{o}$, $66 ^{o}$ 
and $90 ^{o}$ (force along the step).
The case in which the particles move on a  substrate which is free of steps 
( with the force along the x-axis) is also studied to make a comparison
 with the simulations that were done with the step present. 

\begin{figure}
\centerline{
\vbox{ \hbox{\epsfxsize=4.5cm \epsfbox{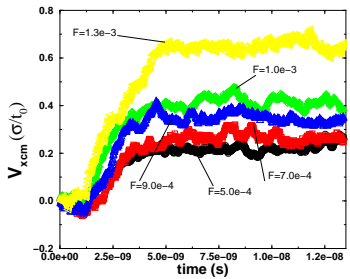} }
       \vspace*{1.0cm}
        } 
}
\caption{The velocity is shown as a function of time 
for the various values of the applied force for the case of 
no step present. The driving force is applied at $0^{o}$ with 
the $x$ axis.}
\label{velocity}
\end{figure}

\begin{figure}
\centerline{
\vbox{ \hbox{\epsfxsize=4.5cm \epsfbox{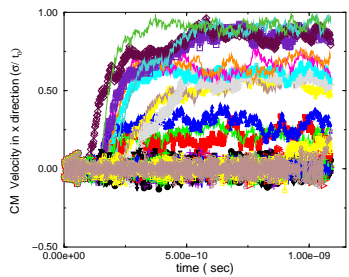} }
       \vspace*{1.0cm}
        } 
}
\caption{Velocity of the center of mass in the $x$ direction 
as a function of time, 
for various values of an applied force, for the case 
of a step present. In this case the driving force is applied at $0^{o}$ 
with the $x$ axis.}
\label{velostep}
\end{figure}

\begin{figure}
\centerline{
\epsfxsize=4.5cm
\epsfbox{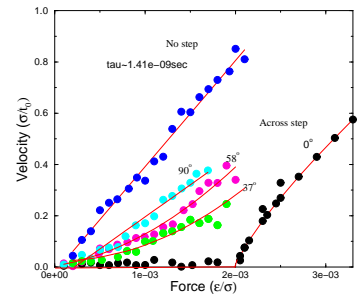}}
\vspace{.5cm}
\caption{Variation of the velocity 
with F at $77.4^{o}$ kelvin.}
\label{depinning}
\end{figure}

In Fig. \ref{velocity}, the velocity is shown as a function of time 
for the various values 
of the applied force for the case of a no step present, 
and in Fig. \ref{velostep} for the case of a step present. 
The velocity at the top of the plateau is the steady-state velocity, 
which we will later plot versus the applied force $F$. 
In order to diminish thermal fluctuations of the velocity vs. time curves, 
the curves shown in Figs. \ref{velocity} and \ref{velostep} are obtained by 
making averages of several runs. The no-step case was 
obtained averaging 5 runs and the step cases, by averaging 3 runs each.
The steady-state velocity is plotted as a function of 
$F$ in Fig. \ref{depinning}. We see that 
the magnitudes of the velocities depend on the orientation of $F$. 
The velocity is smallest at 0$^o$ 
(i.e. when the particles move almost perpendicular to the step).
It gets larger  when the angle increases, taking on its maximum
value for an angle of $90^{o}$ (when particles move along the step).

\begin{figure}
\centerline{
\vbox{ \hbox{\epsfxsize=4.5cm \epsfbox{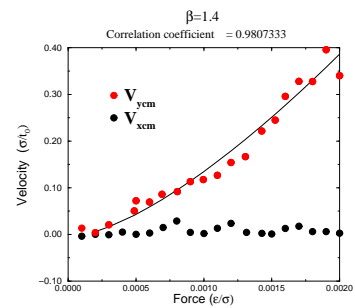}}
        } 
}
\caption{Variation of $v_x$ and $v_y$ 
with F for $58^{o}$ for $2.025$ $meV$ 
of corrugation.} 
\label{sepa}
\end{figure}

When the particles move perpendicular to the step (i.e., in the 
$x$ direction), their motion in $x$ is pinned. For angles of incidence 
$0<\theta<90$ the pinning transition appears to be ``softened''
or ``smeared out.''  It is thus of interest  to plot the $x$ and $y$  
components
of velocity  vs. the applied force (i.e. $V_{x,y}^{cm}$ vs. $F$)
in order to gain insight into what the atoms do. 
In Fig. \ref{sepa},  the $x$ and $y$ components of the velocity are
shown  separately for $58^{o}$  for a 2.025 meV corrugation potential 
strength. We see evidence of 
pinning 
in $x$ but not 
in $y$. (The exponent $\beta$
in this case is $\beta=1.4.$)

\noindent
We also made some runs with a corrugation strength of 3.56 meV. 
The film
moves in the same fashion as for $2.025 meV$ corrugation, but of course 
with much smaller velocities.

Two cases clearly show linearity between velocity and F: no  step 
and $90^{o}$ (along the step). In addition, for angles between 0$^o$ 
and 90$^o$ 
v shown in Fig. \ref{depinning} is linear in F to a good approximation 
for v less than 
0.1$\sigma/t_0$. This is one possible explanation for why such viscous 
friction is found in experiment.

\subsection{Stick Slip and Creep}

\begin{figure}
\centerline{
\vbox{ \hbox{\epsfxsize=4.5cm \epsfbox{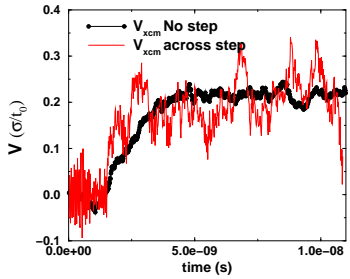}}
        } 
}
\caption{Comparison
of velocity profiles for the step and no step case. Fluctuations 
in $V_{cm}$ are noticeably 
larger when there is a step than in the absence of it.}
\label{compara_velo}
\end{figure}

The results reported in Fig. \ref{depinning} 
are obtained by averaging  
the $V_{cm}$ versus time data for a few different runs for each value 
of the applied force. These averages are performed in 
order to suppress the large fluctuations in the data  found when a 
step is present. In order to illustrate this, $V_{cm}$ 
is plotted as a function of time, without doing any 
averaging, for one value of the applied force larger than the critical 
force  in Fig. \ref{compara_velo}. 
For comparison, a plot of 
$V_{cm}$ versus time for the case of no step present is also 
made on the same graph 
for an applied force  chosen so that $V_{cm}$ has about the same 
value at each time. In this graph the force, for the case in which 
there is a step, has a value  
$F=1.9$ $10^{-3}$ $\epsilon/\sigma$, slightly above threshold.   
\noindent 
It is clear from this figure that the the fluctuations in $V_{cm}$ 
are noticeably 
larger when there is a step than in the absence of a step. It is likely 
that these large fluctuations are due to stick-slip motion, which occurs 
for a system which would be pinned at smaller values of the applied \
critical force.

\begin{figure}
\centerline{
\vbox{ \hbox{\epsfxsize=4.5cm \epsfbox{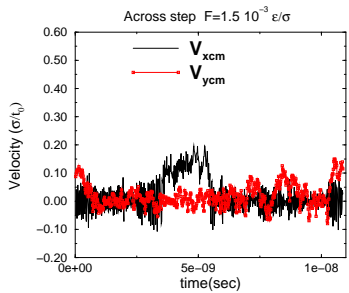}}
        } 
}
\caption{Typical plot of $V_{cm}$  vs time for $F < F_{c}$ 
 The $x$ component (perpendicular to the step) 
and the $y$-component of $V_{cm}$ are shown separately.  
  $V_{cm,x}$ becomes thermally activated  for a short period 
of time and 
then becomes pinned again. Such behavior is 
not found in $V_{cm,y}$}
\label{velo8}
\end{figure}

\begin{figure}
\centerline{
\vbox{ \hbox{\epsfxsize=4.5cm \epsfbox{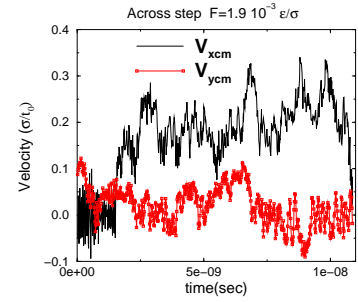}}
        } 
}
\caption{$V_{cm}$  vs. time for $F > F_{c}$ across the step.}
\label{velo10}
\end{figure}

\begin{figure}
\centerline{
\vbox{ \hbox{\epsfxsize=4.5cm \epsfbox{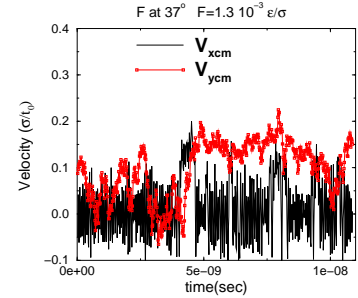}}
        } 
}
\caption{The applied 
force is at an angle of  $37$ degrees with the normal 
to the step. We can  see the thermally 
activated behavior of $V_{cm,x}$ for a couple of short time intervals. 
In contrast, $V_{cm,y}$ appears to saturate at a positive value 
(because the 
film is not pinned in the $y$-direction). Thermally activated motion 
of the 
type that we see here is the type of behavior that would lead to creep 
of a macroscopic film.}
\label{velo11_37}
\end{figure}

     In Fig. \ref{velo8}, a typical plot of $V_{cm}$ is shown for $F$ 
normal to the  step 
below its  threshold value. The $x$ component (normal to the step) 
and the $y$-component of $V_{cm}$ are shown separately.  
  $V_{cm,x}$ becomes thermally activated  for a short period 
of time and 
then becomes pinned again. Such behavior is not found in $V_{cm,y}$. 
The behavior for values of the force slightly above
threshold is shown in Fig. \ref{velo10}. 
In  Fig. \ref{velo11_37} a similar plot is shown 
for the applied 
force at an angle of  $37$ degrees with the normal 
to the step. We can again see the thermally 
activated behavior of $V_{cm,x}$ for a couple of short time intervals. 
In contrast, $V_{cm,y}$ appears to saturate at a positive value 
(because the 
film is not pinned in the $y$-direction). Thermally activated motion 
of the 
type that we see here is the type of behavior that would lead to creep 
of a macroscopic film.

\subsection{Film Rotated Relative to the Step}

\begin{figure}
\centerline{
\vbox{ \hbox{\epsfxsize=4.5cm \epsfbox{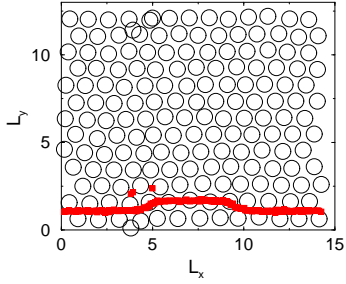}}
        } 
}
\caption{The atomic positions (looking down on the substrate) are shown for 
173 atoms (171 atoms adsorbed on the substrate and two remain above the 
substrate). A side view of the step is also shown at the bottom of the 
figure as a guide to the eye. This film is rotated slightly with respect to 
the walls of the box, and hence with respect to the step.}
\label{rotated}
\end{figure}

In the runs discussed so far, the crystallographic axes of the film were 
lined up with the step. This is not always true for this system. 
For example, we also did a run for a 173 atom film (a more compressed 
monolayer film) for the same substrate containing a step. In this case, 
the film in equilibrium   APPEARED  rotated at an angle with respect 
to the step, 
as seen in Fig. \ref{rotated}. In this case,
 the film attained a steady-state velocity of 0.2$\sigma/t_0$ when 
F=0.0015$\epsilon/\sigma$ was applied normal to the step. As can be seen in 
Fig. \ref{depinning}, the 163 atom film, which was not rotated with 
respect to the step, 
was pinned for F of that magnitude. As we shall see in the next section, 
this is consistent with the predictions of lowest order perturbation theory.

\subsection{Results of the Simulations done with Point Defects Present}

\begin{figure}
\centerline{
\vbox{ \hbox{\epsfxsize=4.5cm \epsfbox{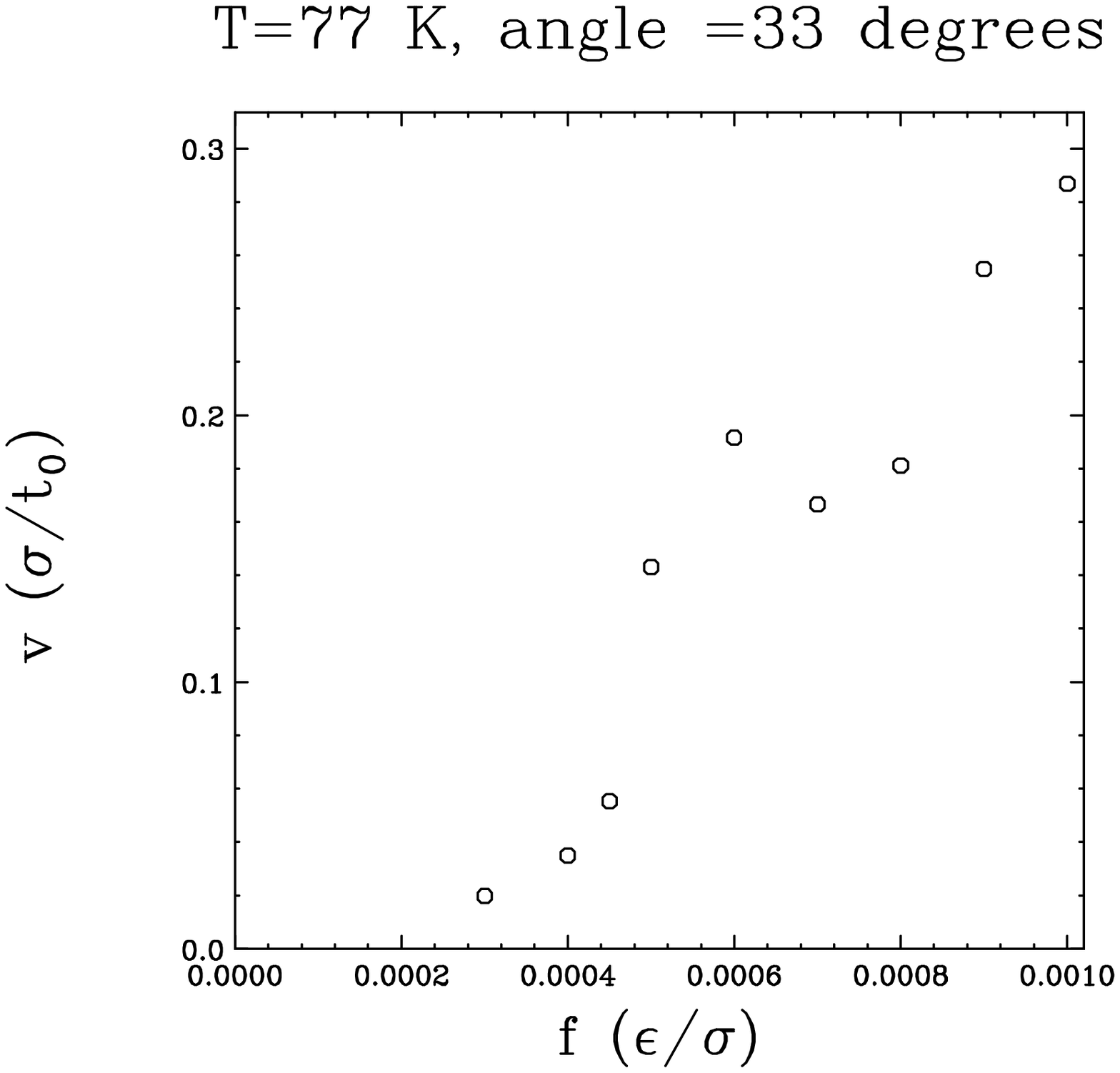}
                \hspace*{0.5cm} \epsfxsize= 4.5cm  \epsfbox{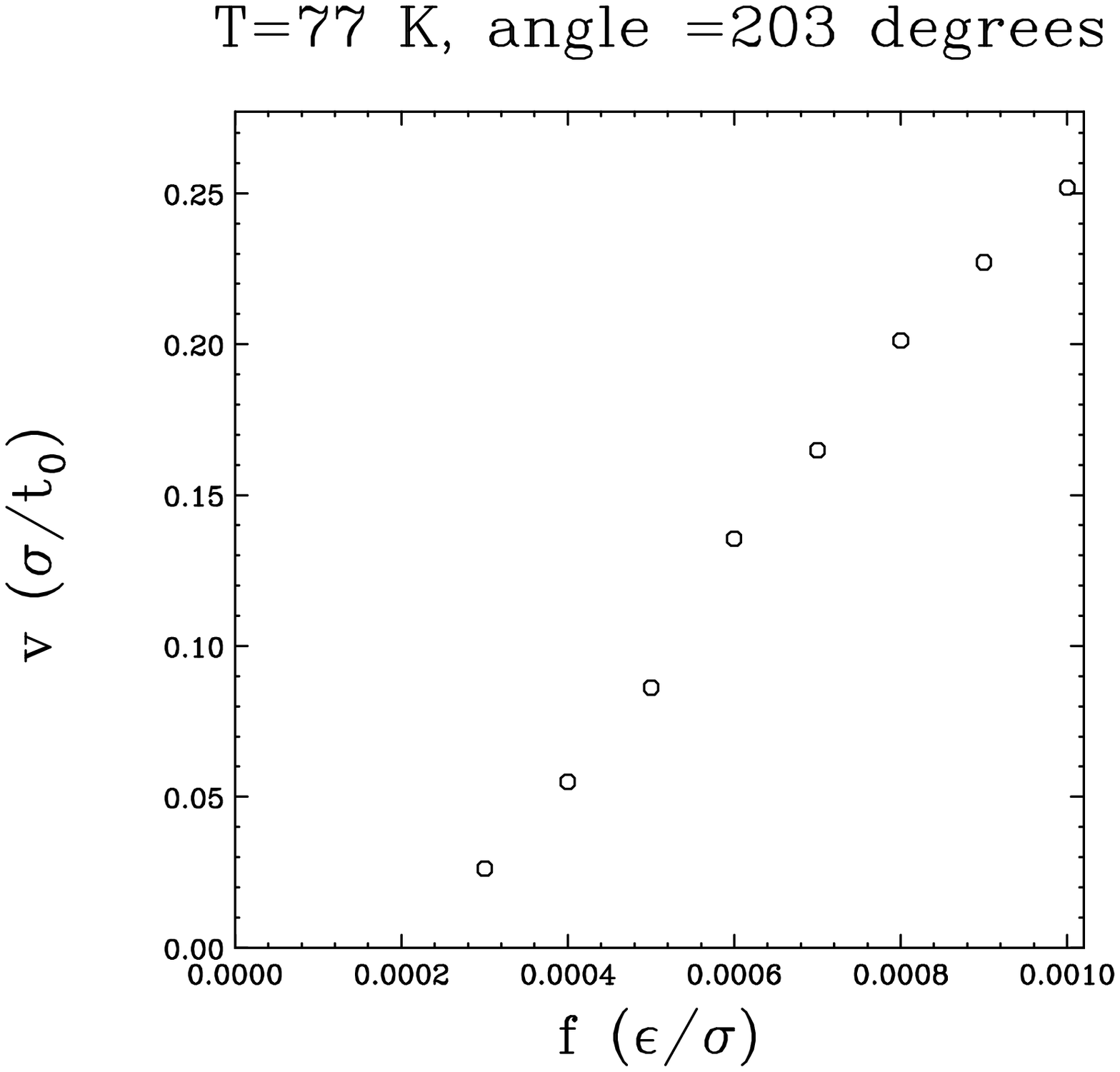} }
       \vspace*{1.0cm}
        } 
}
\centerline{
\vbox{ \hbox{\epsfxsize=4.5cm \epsfbox{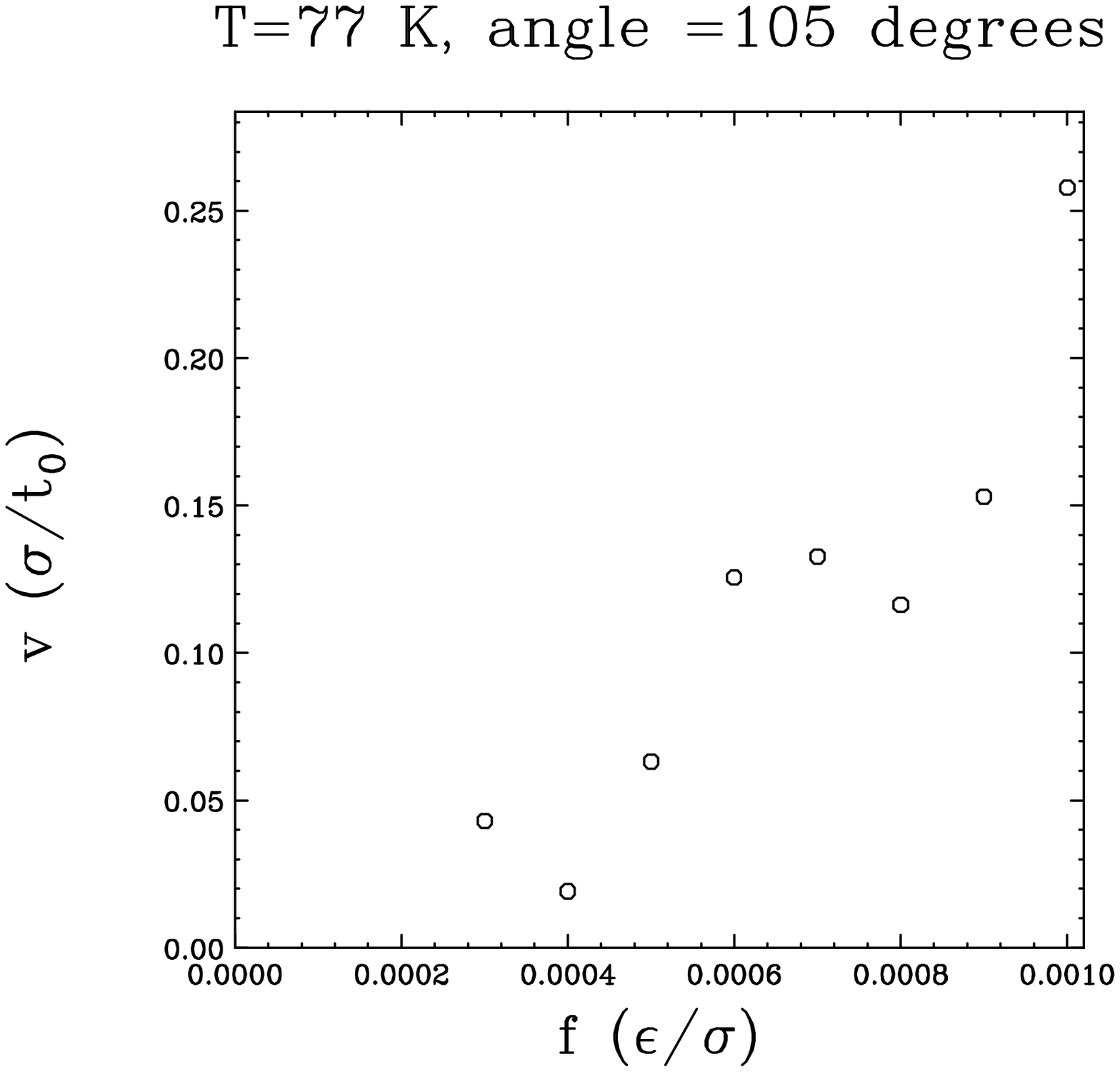}
                \hspace*{0.5cm} \epsfxsize= 4.5cm  \epsfbox{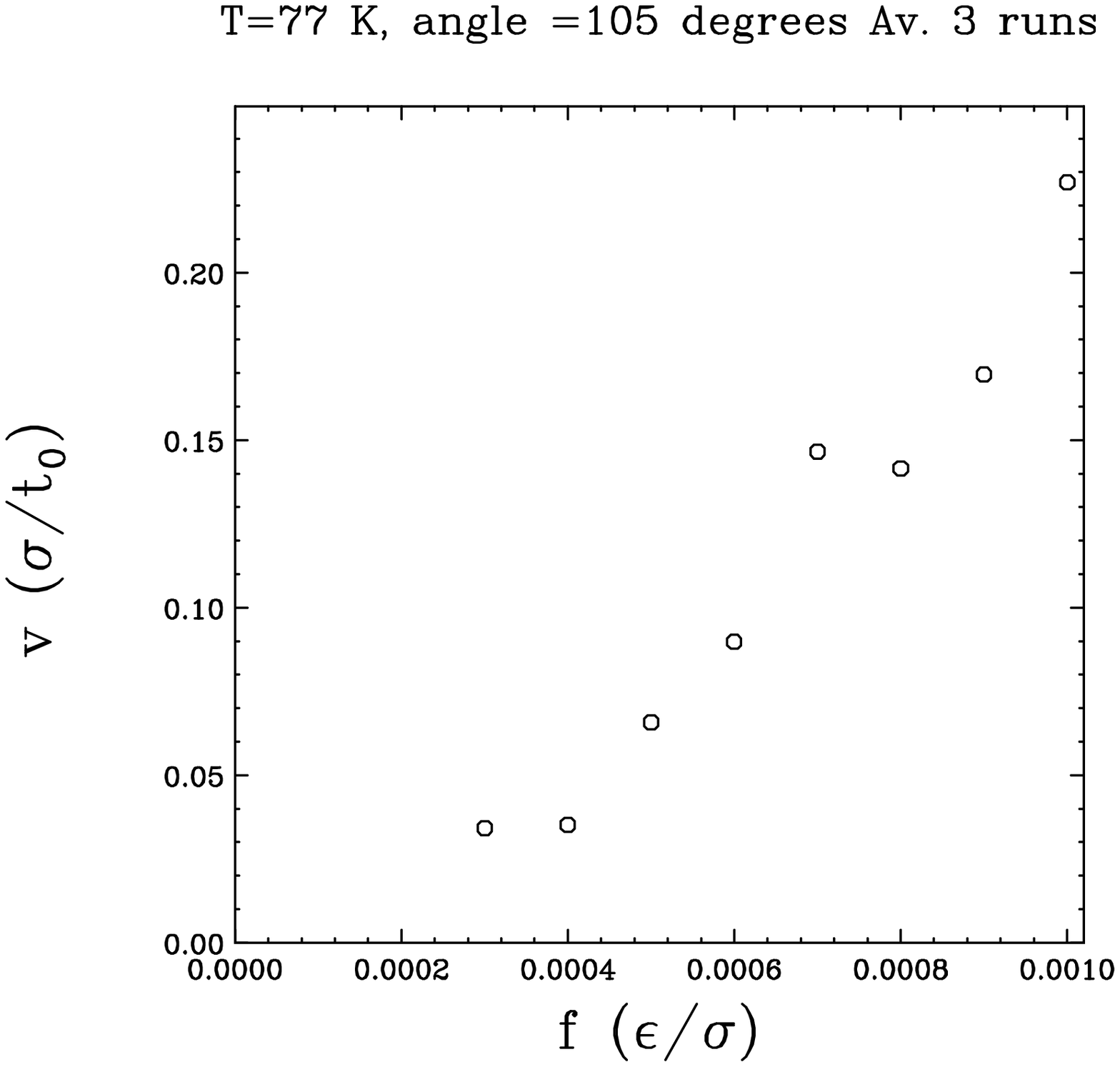} }
       \vspace*{1.0cm}
       }
}
\caption{The velocity in units of $\sigma/t_0$ versus applied force f in 
units of $\epsilon/\sigma$ is shown for a lattice containing a random array 
of point defects with a 5 percent concentration, the force f applied at 33, 105 
and 203 degrees with respect to $L_x$.} An average of three runs is also shown 
for one case.
\label{point1}
\end{figure}

\begin{figure}
\centerline{
\epsfxsize=4.5cm \epsfbox{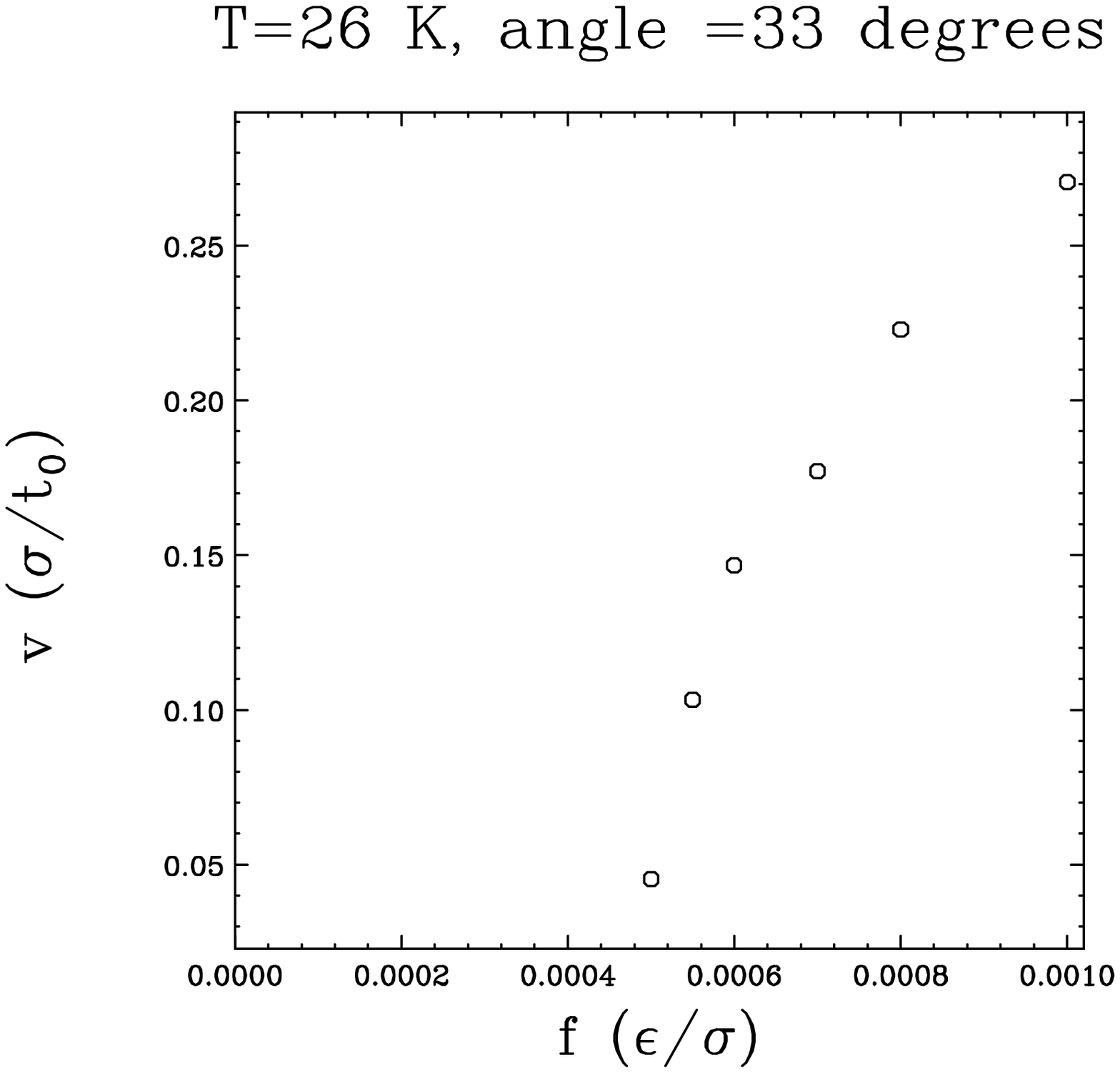}}
       \vspace*{1.0cm}
\caption{The velocity in units of $\sigma/t_0$ versus applied force f in 
units of $\epsilon/\sigma$ is shown for a lattice containing a random array 
of point defects with a 5 percent concentration, with the f making a 33 degree 
angle with $L_x$. This run was done at a temperature of $T=26\,^oK$ .}
\label{point2}
\end{figure}

In these simulations $V_0$ in Eq. \ref{imppot} was chosen to be equal to 
0.1 $\epsilon$, for which each defect potential has a maximum value 
approximately equal to the corrugation potential well depth, and the 
concentration c was chosen to be equal to $0.05$. Again, the system was 
first equilibrated for at least 600,000 iterations, and then an external 
force was applied. Simulations were done with the external force making angles 
of 33, 105 and 203 degrees with respect to the side $L_x$ of the box. 
The results are shown in Fig. \ref{point1}. Each point represents the average 
of the component of the center of mass velocity along the applied force 
over a time interval equal to at least $10^4t_0$ (i.e. $2\times 10^6$ 
iterations or about $3\times 10^{-8}$ s.) in order to 
average out the fluctuations. 
An average of three runs is also shown for one of the cases. Since it did 
not look qualitatively different than the single runs, it is clear that 
single runs are adequate. For all three directions of the applied 
force the film appeared to become pinned below a critical force of about 
$2.10^{-4}$ $\epsilon/\sigma$, even though the detailed shape of the velocity 
versus force curves might have differed slightly. A run was also done at 
a temperature of $26$ $K$ in order to determine whether the motion of the film 
might have been partly thermally activated. The results are shown in 
Fig. \ref{point2}.   
 It is seen that the critical field is now about $4.10^{-4}$ $\epsilon/
\sigma$, and the curve appears to exhibit a sharp depinning transition, 
whereas the runs done at $T=77.4\,^oK $  showed what looked more 
like a rounded transition. These results imply that some thermal activation
 of the atoms out of the impurity potential wells is taking place. 
It is quite unlikely,
however, that such thermal activation, which is responsible for creep, 
will give such a large contribution for a macroscopic film as it does 
for the small films used in the simulations. 

The pinning force of $2.10^{-4}$ $\epsilon/\sigma$ found at $T=77.4\,^oK$ 
 is equal to about
$1.5\times 10^{-11}dyn$ per atom. 

One might be tempted to think that the fact that point defects do not seem 
to have much of an influence on QCM friction measurements can be explained 
by relatively small value that we have found for the pinning force. We do 
not believe that this is correct, however, because 
this force is still much larger than the effective inertial force due to 
the oscillations of the QCM's substrate  which is equal to 
$m\omega^2 A\approx 10^{-14}dyn$, where $m$ is a film atom mass ($\approx 
10^{-22} dyn$, $\omega\approx 10^7 s^{-1}$ and the amplitude of the 
substrate oscillations $A\approx 10^2 A^o$. 


\section{Perturbation Theoretic Treatment}

The force of friction acting on a 
thin film as it slides over a substrate containing defects was recently 
studied using 
lowest order perturbation theory in the substrate potential \cite{soko-toma}. 
It was shown there that for most orientations of the film and most 
directions of sliding 
over the substrate, the force of friction is viscous (i.e., proportional
to the sliding velocity), in agreement with what has to date been 
observed in QCM
friction  experiments \cite{krim1}. The simulations that were 
discussed in the last section of a xenon film sliding over 
a silver (111) substrate containing a step show that the film is pinned  
if we apply a force smaller than a critical value 
in a direction normal to the step. When a force larger than 
this critical value is applied, the resulting velocity is a nonlinear 
function of the applied force. 
In this case the film's axes are lined 	up with the step.
For many substrate-adsorbed film systems, the case considered in the 
simulations, 
in which the crystallographic axes of the film are lined up with the 
step, is 
an important case because one expects that it will be 
energetically favorable for the film 's axes to line up with the step locally. 
The reason for expecting this to occur is that adsorbate atoms can lower 
their energies by lying along a step edge, because in this way each of 
these adsorbate atoms will be surrounded on two sides by a substrate atom. 
For the present system, our simulations seem to show, however, that the 
film does not always line up with the step, and when it is not lined up 
(for example, for the 173 atom system discussed at the end of the last 
section), 
the simulations show that the film is not pinned for forces for which 
it would be pinned if it were. (For real surfaces, which have many steps, 
it is expected that because of its stiffness the film in its solid phase 
will typically 
not be able to distort in such a way that its crystallographic axes line up 
locally with each step.) 
In this section we will try to interpret the results of the simulations
 reported here using the perturbation theoretic methods of reference.\cite 
{soko-toma} 

In Ref. \cite{soko-toma}, 
it was shown using perturbation theory that 
for a general 
nonperiodic substrate force, the mean force of 
friction acting on a thin film is given by 
\begin{equation}
\label{3.1} F_{av}v=(mN)^{-1}\sum_{{\bf k},\sigma}
{\gamma\omega^2 |{\bf f}({\bf k})\cdot\hat 
{\bf \epsilon}_{{\bf k},\sigma}|^2\over (\omega^2_{\sigma}({\bf k})
-\omega^2)^2+\gamma^2
\omega^2}, 
\end{equation}
where $\omega=v_x k_x+v_y k_y$, where ${\bf v}$ is the sliding velocity 
of the 
film, ${\bf f}({\bf k})$ is the Fourier transform of the force due to the 
substrate, m is the mass of an atom, N is the number of atoms in the film 
and $\gamma$ is the inverse phonon lifetime. This expression was obtained 
by setting the rate at which the substrate force $\vec f(\vec r-\vec v t)$ 
does work on the film equal to the rate at which the average force of friction 
$F_{av}$ does work on the film, i.e., $F_{av}v.$ 
It was also found in that reference that the 
mean square vibrational displacement of an atom in the film (in the rest 
frame of the film) is given by 
\begin{equation}
\label{3.2} <u^2>=m^{-2}N^{-1}\sum_{{\bf k},\sigma}
{|{\bf f}({\bf k})\cdot\hat 
{\bf \epsilon}_{{\bf k},\sigma}|^2\over (\omega^2_{\sigma}({\bf k})
-\omega^2)^2+\gamma^2
\omega^2}. 
\end{equation}
In Ref. \cite{soko-toma}, it was found that in two or fewer dimensions 
for a substrate 
containing point defects (e.g., vacancies or substitutional impurities), 
$<u^2>$ diverges 
as ${\bf v}$ approaches zero. This signifies that in the zero velocity limit, 
there will always be significant distortion of the film to conform to the 
substrate, no matter how small the substrate force, which implies that 
the film 
will be pinned in place unless a strong enough external force is applied. 

In the $\gamma$ approaches zero limit, Eq. (\ref{3.1}) becomes 
\begin{equation}
\label{3.3} F_{av}=(\pi/m)(a/2\pi)^2\sum_{\sigma}
\int d^d k \omega |{\bf f}({\bf k})|^2 
\delta (\omega^2_{\sigma}({\bf k})-\omega^2). 
\end{equation}
In this limit, a nonzero contribution to $F_{av}$ occurs when the 
argument of the delta function is zero, which occurs when the plane in the 
d+1 dimensional space defined by $\omega$ and the d-components of 
${\bf k}$ whose equation is 
$\omega={\bf v}\cdot{\bf k}$ intersects the phonon dispersion surface 
[whose equation is $\omega=\omega_{\sigma}({\bf k})$], 
if ${\bf f}({\bf r})$ were aperiodic for all directions of ${\bf r}$. 
This is the case when we 
have point defects in the substrate. It was shown in Ref. \cite {soko-toma}
that in this 
case in lowest order perturbation theory for a two dimensional (i.e., 
monolayer) film $F_{av}$ is independent 
of velocity and $<u^2>$ diverges as 1/v as v approaches zero, implying 
pinning of the film at low velocities. In contrast for a three 
dimensional (i.e., thick) sliding film, $F_{av}$ was found to be 
proportional to v and $<u^2>$ 
did not diverge as v became zero, which is consistent with the film not 
being pinned. In appendix C, these ideas are expanded upon by considering 
general order in perturbation theory. It is shown that d=2 is indeed a 
critical dimension for this problem.

Let us now consider a line defect, such as a step or facet boundary. 
If we choose the x and y axes (i.e., the coordinate axes in the plane 
of the 
film) so that the y-axis is along the defect (i.e., along a step or facet 
boundary, which are taken in the present discussion to be straight lines) 
${\bf f}({\bf k})$ will be non-zero 
only if $k_y$ is a multiple of the y-component of one of the reciprocal 
lattice 
vectors of the substrate, since the substrate is periodic in that 
direction if 
the crystallographic axes of the substrate are lined up with the defect and 
quasiperiodic otherwise. As a consequence, the plane whose equation is 
$\omega=v_x k_x+v_y k_y$ gets fragmented into a series of parallel lines 
at values of $k_y$ equal to $Q_y$, the y-component of one of the smaller 
reciprocal lattice vectors of the substrate. For a facet boundary or 
step, ${\bf f}
$ can be written as ${\bf f}({\bf r})=g(x){\bf F}({\bf r})$, 
where ${\bf F}$ is periodic but g is not. (This 
was illustrated for the model for the potential that was used in the 
simulations presented in Ref. \cite {soko-toma}). A reasonable model 
potential for other line defects, such as 
facet boundaries, for example, will have a similar form. This is discussed 
in appendix B. Since we can 
write ${\bf F}$ as ${\bf F}({\bf r})=\sum_{\bf Q}{\bf F}_{\bf Q}e^{i{\bf Q}
\cdot{\bf r}}$, we find that 
\begin{equation}
\label{3.4} {\bf f}({\bf k})=\int d^2 r e^{-i{\bf k}\cdot{\bf r}}
{\bf f}({\bf r})=\sum_{\bf Q}
\delta_{k_y,Q_y}g_{k_x-Q_x}{\bf F}_{\bf Q}, 
\end{equation}
where ${\bf Q}$ denotes a reciprocal lattice vector of the substrate. 
It is only when one of these 
lines in k-space intersects the  phonon dispersion surface that Eq. (\ref{3.3}) 
gives a 
nonzero contribution to $F_{av}$. If this does not occur, we must keep 
$\gamma$ 
nonzero in Eq. (\ref{3.1}). We then find that $F_{av}$ is proportional to 
v\cite{soko-toma}. 

Let us now consider the case in which the crystallographic axes of the film 
are not lined up with the direction of the defect but are close to being so. 
Furthermore, let us keep $\gamma$ nonzero for this discussion, in order 
to examine what happens if a line passes within a phonon linewidth of the 
phonon dispersion surface Then taking the defect to be lined up with the 
y-axis, if we convert the summation over wavevector in Eq. (\ref{3.2}) to an 
integral along a line perpendicular to the step in the usual way, we obtain
\begin{equation}
\label{3.5a} <u^2>/a^2={a\over 2\pi m^2a^2}\sum_{\sigma}\int dk_x 
F(k_x,k_y,G,v)
\end{equation}
where $F(k_x,k_y,G,v)$ is given by 
\begin{eqnarray}
\label{3.5b} 
{|{\bf f}(k_x,k_{y0})\cdot\hat{\bf \epsilon}_{{\bf k},
\sigma}|^2 \over
[v_p^2 (k_x^2+k_{y0}^2)-\omega^2]^2
+\gamma^2 \omega^2},\nonumber\\
\end{eqnarray}
where $\omega=v_y k_{y0}+G_x v_x$ and 
where $k_{y0}$ is the amount that the line, which would have passed 
directly through the point in k-space denoted by the film reciprocal 
lattice vector ${\bf G}$ if the film's 
crystallographic axes were line up with the defect, misses going through 
this point in reciprocal space. On doing the integral over $k_x$ by 
contour integration we find that $<u^2>/a^2$ is equal to 
\begin{eqnarray}
\label{3.6} 
(ma)^{-2}v_p^{-1}({3a\over 2\pi}){\pi 
|{\bf f}\cdot\hat{\bf \epsilon}|^2 \over 2sin (\theta_1/2)(v_p^4 k_{y0}^4
+\gamma^2 G_x^2 v_x^2)^{3/4}},\nonumber\\
\end{eqnarray}
where $|{\bf f}\cdot{\bf \epsilon}|$ is evaluated at the 
wavevector corresponding 
to the pole in the integrand of Eq. (\ref{3.5a}) and 
where $\theta_1=arctan [\gamma v_x G_x/(v_x^2 G_x^2-v_p^2 k_{y0}^2)].$ 
(Note that $sin\theta_1/2$ becomes 1 rather than zero as $v_x$ approaches 
zero, since $\theta_1$ approaches $\pi$ rather than zero. 
>From these results, we see that if $k_{y0}\not=0,$ $<u^2>/a^2$ does not 
diverge as $v_x$ approaches zero, which implies that the film is not pinned. 
Furthermore, if $|{\bf f}\cdot\hat{\bf \epsilon}|$ 
is of the order of $10^{-7}dyn$, $m\approx 10^{-22}g$, $a\approx 10^{-8} cm$, 
$G_y\approx 10^8 cm^{-1}$ and $k_{y0}\approx 10^7 cm^{-1}$ (or about 1/10 of 
the Brillouin zone radius), we estimate that $<u^2>/a^2$ is of the 
order of $10^{-3}.$ Thus, we conclude that for a reasonable value of the 
mis-orientation parameter $k_{y0}$, lowest order perturbation theory 
should be a correct description of the friction, even as v approaches zero. 
It is easily seen from 
Eqs. (\ref{3.1}) and (\ref{3.2}) that for this case, in which $<u^2>/a^2$ 
does not 
diverge as the sliding velocity approaches zero, the force of friction will 
be proportional to the velocity. 
\section{\bf Speculations On the Behavior of Real Interfaces} 
We propose the following  physical 
explanation for why a line defect, 
such as a step, will pin the film if it is oriented along a crystallographic 
axis of the film but will not pin it if it is not lined up: When the 
line defect
 is lined up with a crystal axis, a line of atoms running across the 
width of 
the film will pass over the defect all at the same time as the film 
slides over 
the defect. Consequently, the force exerted by the defect on the film will 
be an extensive quantity, in the sense that it will scale with the width
of the film. In contrast, when the defect is not lined up with an axis, 
atoms 
in the film will pass over the defect one at a time. The defect will act on 
one atom at a time with a force that does not scale with the width of the 
film. Thus, if we consider a situation in which the velocity of the film 
is studied as a function of a force applied to each atom in the film, 
when the 
film's axes and the defect are not lined up, it will clearly be much easier 
for the applied force to overcome any nonextensive pinning force due to the 
defect than an extensive one. A real substrate surface will contain a 
finite density of line defects along a line drawn along the 
sliding velocity. These defects will most likely not run across the width 
of the film, and they certainly will not be straight over that distance. 
Then, 
although the pinning force due to one of these line defects will not be 
extensive, the force due to a finite density of them will.

One way to understand the case of many finite length line defects, which 
are not perfectly 
straight, as occur on real surfaces using the calculation presented above 
for a single straight line defect is that if the defect is not a straight 
line over an infinitely 
long distance, the Fourier transform of the substrate force in 
Eq. (\ref{3.4}) 
will 
no longer contain a $\delta_{k_y,Q_y}$ factor (since a kroniker delta only 
occurs if ${\bf f}$ is periodic over an infinite distance). Rather the 
Kroniker delta 
will be replaced by a function of $k_y$ of finite width of the order of 
$2\pi/\ell$, where $\ell$ is the length of a straight section of the 
defect, peaked around 
the values of $Q_y$. Consequently, we can now think of the lines in k-space, 
defined by $k_y=Q_y$,
which must intersect the phonon dispersion surface in order for the delta 
function in Eq. (\ref{3.3}) to be satisfied, as being broadened. For real 
surfaces, which contain many finite length straight sections of line defects 
separated by defect-free regions, ${\bf f}({\bf k})$ will consist of the 
following contributions: defect-free regions which will give a peak in 
${\bf f}$ at each 
reciprocal lattice vector of the film, point defects, which will give a 
contribution which is not peaked in k-space and 
straight sections of line defects, which
 will each give a contribution that is peaked at values of the component of 
${\bf k}$ along the defect equal to 
components of a reciprocal lattice vector of the substrate 
if we move in k-space in a direction 
parallel to the defect. How sharply it is peaked will depend on the 
length of 
the straight section of the defect. In the direction perpendicular to 
the defect, the contribution to ${\bf f}({\bf k})$ will not be peaked. The 
magnitude of each of these contributions to ${\bf f}({\bf k})$ will be 
proportional to the number of lattice sites in each of the above elements. 
Thus, 
when we take the square of ${\bf f}({\bf k})$, which enters Eqs. (\ref{3.1})
and (\ref{3.2}), we will thus obtain the following contributions: 1. a Bragg 
peak at each substrate reciprocal lattice vector, 2. a contribution which 
is not peaked in k (from the point defects), and 3. a series of 
"mountain ridges" in k-space directed in various directions. These are the 
lines in k-space that we found for a single line defect except that each one 
now has a width of the order of $2\pi$ divided by the length of the 
straight section of the line defect that gives this contribution. If the 
widths of the "mountain ridges" 
are fairly narrow, the ridges will generally miss passing through a dip 
in the phonon dispersion surface, which as discussed in the last section, is 
the criterion for the delta function in Eq. (\ref{3.3}) being satisfied). 
Then, these will give mainly a viscous contribution to the friction, 
following 
the discussion in the last section. On the other hand, if the straight 
sections 
of the line defects are relatively short, resulting in broad ridges, the 
ridges will generally intersect the phonon dispersion surfaces, most likely 
leading to "dry" (i.e., velocity independent) friction. The point defects 
will always contribute "dry friction."We know from 
the discussion in the last section that the Bragg peaks will give viscous 
friction in perturbation theory, as long as their widths are not too large. 
More numerical studies of this kind, which use perturbation theory to study 
the effects of realistic models for defected surfaces will be presented in 
future work.

\section{\bf Conclusions}

Simulations and perturbation theory calculations done for a monolayer 
film sliding on a substrate 
containing a step defect gave approximately viscous friction for all 
sliding directions other than perpendicular to the step. 
The point defects in the simulations were found to pin the film below a 
critical applied force  For defects of strength comparable to the strength 
of the corrugation potential (which were used in the present simulations), 
however, the pinning force was only about $10^{-11}dyn.$ This force is 
still greater than the inertial effective force resulting from the oscillations 
of the microbalance ($m\omega^2 A$, where $m\approx 10^{-22}g$ is the adsorbate 
atom mass and $\omega\approx 10^7s^{-1}$ and $A\approx 10^{-6}cm$ are the 
microbalance frequency and amplitude, respectively), which is about 
$10^{-14}dyn$. The fact that the observed force of friction generally 
seems to be viscous in these experiments seems to imply that either the defects 
are much weaker or have a concentration much lower than the 5 percent 
concentration used in the present simulations. 

\acknowledgments 

We wish to thank the United States Department of Energy's Office of Basic 
Energy Sciences for their support in the form of Grant NO. 
 DE-FG02-96ER45585. This work benefited from the allocation of time at the 
Northeastern University High-Performance Computing Center (NU-HPCC).
We would like to thank Professor H.E. Stanley for his aid. One of us (M.S.T) 
would like to thank H. A. Makse for useful discussions.

\appendix
\section {Effects  of Periodic Boundary Conditions}

     We will now use perturbation theory\cite 
{soko-toma} to estimate the
effect of using periodic boundary conditions in simulations of films 
sliding 
over substrates containing steps and substrates containing point 
defects (e.g.,
vacancies, substitutional impurities or add-atoms). What we will do is to 
calculate the force of friction for a system with periodic boundary 
conditions 
and compare our results 
to the results of the calculation of the force of friction for an infinite 
system.  
For the case that we consider, motion takes place perpendicular to the step. 
Since the finite system with periodic boundary conditions is periodic 
with the length of the box, 
the driving force due to the substrate (in which we are doing 
perturbation theory) is periodic. As a result, the sum over wave-vector 
in the 
expression for the force of friction, i.e., Eq. (5) in 
Ref. \cite {soko-toma}, becomes Eq. (12) in that reference, which gives the 
force of friction when the substrate is perfectly periodic 
with reciprocal lattice vectors with components $G_x=2\pi n_1/L$ and 
$G_y=2\pi n_2/L$, where L is the length of the box (the reciprocal lattice 
vectors for a substrate which is periodic only because of the use of periodic 
boundary conditions). Then the important 
parameters 
are R, the ratio of $2\pi/L$ to the reciprocal lattice vector of the 
substrate,
the phonon mode width $\gamma$, and the sliding velocity v. 

\begin{figure}
\centerline{
\vbox{ \hbox{\epsfxsize=4.5cm \epsfbox{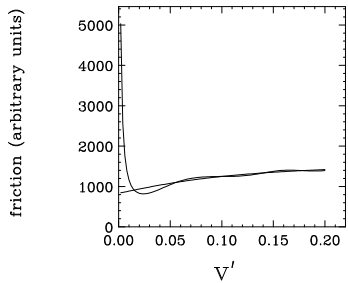}}
       \vspace*{1.0cm}
        } 
}
\caption{A comparison of Eq. (\ref{A2a}) (the curve that is higher near 
v'=0) and (\ref{A2b}), which are 
proportional to the force of friction for a finite and an infinite 
system, respectively, with periodic boundary conditions.}
\label{pointdef}
\end{figure}

To be explicit, we do the sum over wavevector in Eq. (5) of Ref. 
\cite {soko-toma}. When periodic boundary conditions 
are used with a finite size box, this is a sum over discrete values of k
(i.e., $ k_x=n_1 2\pi/L$ and $k_y=n_2 2\pi/L$, where L is the box 
length). We 
make the following assumptions, valid for small sliding velocity: We 
take the 
sliding velocity to be along the x-axis. We assume that the major 
contribution 
to this summation comes from the intersection of the plane 
$\omega=v k_x$ and 
$\omega=\omega_{\sigma}({\bf k})$. In the small v limit for point 
defects, 
Eq. (5) of Ref. \cite {soko-toma} can be approximated by

\begin{equation}
\label{A1} F_{av}\propto \sum_{{\bf k}}{\gamma G_x^2 v 
|{\bf f}\cdot{\bf \epsilon}|^2\over
(v_p^2 k^2-v^2 G_x^2)^2+\gamma^2\omega^2},
\end{equation}
where $G_x$ is the x-component of one of the smallest reciprocal lattice 
vectors of the film. The k-dependence of 
${\bf f}({\bf k})\cdot{\bf \epsilon}_{{\bf k},\sigma}$ is neglected and 
the origin 
of k-space has been moved to the point ${\bf G}$, one of the smallest 
reciprocal 
lattice vectors of the film. Then, we must evaluate the 
summation 
\begin{equation}
\label{A2a} \sum_{n_1,n_2}{\Gamma v' \over [(n_1^2+n_2^2)R^2-v'^2]^2+
(\Gamma v')^2},
\end{equation}
where $R=2\pi/(LG_x)$, $v'=v/v_p$ and $\Gamma=\gamma/v_p G_x$. 
This summation is evaluated for R=1/13, $\Gamma=0.1$ and v' ranging from 0 to 
0.2,
which are parameters appropriate to the simulations. In the infinite L limit,
the summation over $n_1$ and $n_2$ can be approximated by the integral
\begin{equation}
\label{A2} R^{-2}\int dq_x dq_y {\Gamma v' \over (q^2-v'^2)^2+(\Gamma v')^2},
\end{equation}
where $q_x=n_1 R$ and $q_2=n_2 R$. The result of doing this integral is
\begin{equation}
\label{A2b} (R^{-1})[\pi/2+arc tan(v'/\Gamma)].
\end{equation}
Plots of Eqs. (\ref{A2a}) and (\ref{A2b}) are shown in Fig. \ref{pointdef}. 
As can be seen, the integral 
(which describes an infinite system) and the summation (which describes a 
finite system with periodic boundary conditions) agree with each other for 
v' greater than 0.02. Thus, for velocities 
greater than this value, we are justified in assuming that the simulations 
will 
be a good description of a film on an infinite nonperiodic substrate. 

\begin{figure}
\centerline{
\vbox{ \hbox{\epsfxsize=4.5cm \epsfbox{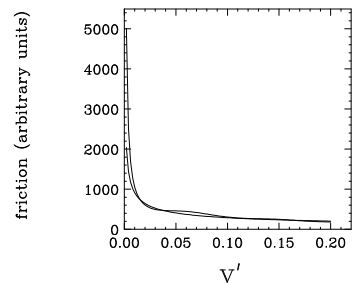}}
       \vspace*{1.0cm}
        } 
}
\caption{A comparison of Eq. (\ref{A3a}) (the curve that is higher near 
v'=0) and (\ref{A3b}), which are 
proportional to the force of friction for a finite and an infinite 
system, respectively, with periodic boundary conditions.}
\label{stepdef}
\end{figure}

For the case of a straight step, the film is periodic or quasiperiodic 
along the step with a repeat distance or almost repeat distance comparable 
to a lattice constant. Thus, if we take the step to be along the y-axis 
and consider the sliding velocity to be in the x-direction,  
the expression for $F_{av}$ involves an integral only over the 
x-component of 
the wavevector for an infinite system. For a finite system with periodic 
boundary conditions, the summation and integral corresponding to these 
quantities are respectively 
\begin{equation}
\label{A3a}\sum_{n} {\Gamma v' \over (n^2 R^2-v'^2)^2+(\Gamma v')^2}
\end{equation}
and
\begin{equation}
\label{A3b}R^{-1}\int dq {\Gamma v' \over (q^2-v'^2)^2+(\Gamma v')^2},
\end{equation}
where q=nR. This integral can easily be performed by contour 
integration to 
give $(\pi/(2\gamma)^{1/2}(v')^{-1/2}$. This result together with the 
summation 
over n are shown in Fig. \ref {stepdef}. We again see that the finite 
system with 
periodic boundary conditions is an accurate representation of the infinite 
system for v' greater than 0.02.

\section{The Substrate Force with a Line Defect Present}

A good model for the corrugation potential when there is a step present 
in the 
crystal which is parallel to the y-axis is
\begin{equation}
\label{B1} v_1 (z-g(x))\sum_{{\bf Q}_1} e^{i{\bf Q}_1\cdot 
({\bf r}-{\bf d}(x))},
\end{equation}
where $g(x)$ is a Fermi function $z_1/[e^{\beta (x-\delta)}+1]$ times 
$z_1$ (where $z_1$ is the height of the step and $\beta$ is the reciprocal 
width of the step edge, which is of the order of the  reciprocal of 
an atomic spacing), ${\bf Q}_1$ denotes the 6 
smallest reciprocal lattice vectors of the substrate and 
${\bf d}(x)$ is the product of a vector giving projection of the difference 
between the equilibrium positions of the film atoms in the 
first and second atomic layers parallel to the x-y plane and the above Fermi 
function. Then the x and y 
components of the force $f_{x/y}$ due to the corrugation potential are 
given by
\begin{eqnarray}
\label{B2}f_x=-\frac{(dV_1 (z_0)}{dz})\sum_{{\bf Q}_1}(\frac{d}{dx})[g(x) 
e^{-i{\bf Q}_1\cdot{\bf d}(x)}]e^{i{\bf Q}_1\cdot{\bf r}}+\nonumber\\
i\sum_{{\bf Q}_1}Q_{1x}
e^{i{\bf Q}_1\cdot ({\bf r}-{\bf d}(x))}
\end{eqnarray}
and 
\begin{equation}
\label{B3}f_y=V_1 (z_0)\sum_{{\bf Q}_1}iQ_{1y}e^{i{\bf Q}_1\cdot 
({\bf r}-{\bf d}(x))},
\end{equation}
where $z_0$ is the value of z at which the attractive potential $V_0 (z)$ 
is minimum. Then it is easily found that each term in $f_x$ and $f_y$ will 
have the form of Eq. (\ref{3.4}).

Now, let us consider another type of line defect, a facet boundary. 
In going from one facet to its neighbor, the crystal surface rotates 
in the 
x-z plane for a facet boundary running along the y-axis. Then a simple 
phenomenological model for a facet boundary, along the lines of the model 
discussed above for the step, is obtained by allowing the rotation angle 
to depend on x. Then, we may write 
\begin{equation}
\label{B4}x'=x cos\theta (x) -z sin\theta (x) \approx x-z\theta (x),
\end{equation}
\begin{equation}
\label{B5}z=z cos\theta (x)+xsin\theta (x)\approx z+x\theta (x),
\end{equation}
since the rotation angle $\theta (x)$ is small. 
Then we obtain 
\begin{eqnarray}
\label{B6}f_x =\sum_{{\bf Q}_1}[-dV_1/dz'|_{z'=z_0}(\theta (x)+x\theta' 
(x))+
\nonumber\\
iQ_{1x}V_1 (z_0) (1-z\theta' (x))]e^{-Q_{1x}z_0\theta (x)}
e^{{\bf Q}_1\cdot{\bf r}},
\end{eqnarray}
whose Fourier transform will clearly be proportional to 
$\delta_{k_y,Q_{1y}}$, 
as is Eq. (\ref{3.4}). The fundamental difference between these two line 
defects is that whereas dg/dx, which determines the strength of the 
nonperiodic 
part of the substrate force (due to the presence of the step) is of order 
1, $\theta$, which determines the strength of the nonperiodic part of the 
substrate force due to the facet boundary, is much smaller, since $\theta$ 
is the angle between two neighboring facets, which is likely to be only 
about a tenth of a radian. Thus the force due to a facet boundary is an 
order of magnitude smaller than that due to a step. 

\section{Behavior of a General Term in Perturbation Theory}

\begin{figure}
\centerline{
\vbox{ \hbox{\epsfxsize=4.5cm \epsfbox{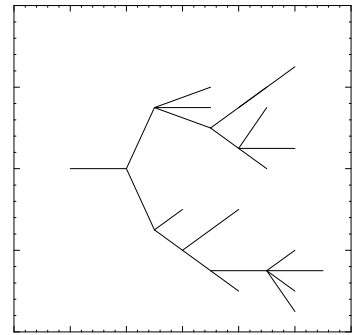}}
       \vspace*{1.0cm}
        } 
}
\caption{Diagramatic representation of the perturbation expansion for 
${\bf u}_j$ in powers of f. The lines and vertices are defined in the 
text.}
\label{tree}
\end{figure}

Arguments similar to those used by Fisher for a similar 
problem, that of a charge density wave moving in a crystal lattice in 
an applied electric field\cite {fisher}, can be used to examine the 
behavior of 
the $n^{th}$ order term in perturbation theory in the substrate potential 
as the sliding velocity approaches zero. (The charge density wave problem 
differs from the present problem, however, in that in the charge density 
wave 
problem, the motion is overdamped, whereas in the present problem it is 
not.)
To accomplish this, 
let us consider the average rate at which the substrate force does work on 
the film by generating phonons, as the film slides over it (which can be set 
equal to $F_{av}v$ where $F_{av}$ is the mean force of friction and v 
is the 
sliding velocity). From previous 
work on this problem this is given by\cite {sokoloff}
\begin{equation}
\label{C1}F_{av}v=(T^{-1})\int dt\sum_{j}
{\bf f}({\bf R}_j+{\bf u}_j-{\bf v}t)\cdot
\dot{{\bf u}}_j (t), 
\end{equation}
where 
\begin{eqnarray}
\label{C2}{\bf u}_j (t)=\sum_{j'}\int dt' 
{\bf G}({\bf R}_j-{\bf R}_{j'},t-t')\cdot \nonumber\\
{\bf f}({\bf R}_{j'}+{\bf u}_{j'}(t')-{\bf v} t'), 
\end{eqnarray}
where the Green's function is given by 
\begin{equation}
\label{C3}{\bf G}({\bf R}_j-{\bf R}_{j'},t-t')=\sum_{{\bf k},\sigma}
\int d\omega
{e^{i{\bf k}\cdot({\bf R}_j-{\bf R}_{j'})}e^{\omega (t-t')} \over
\omega^2_{\sigma} ({\bf k})-\omega^2+i\gamma\omega}. 
\end{equation}
The perturbation series in the substrate potential corrugation strength 
f can be generated by expanding ${\bf f}({\bf R}_{j'}+{\bf u}_{j'}-
{\bf v} t')$ 
on the right hand side of Eq. (\ref{C2}) in a Taylor series in 
${\bf u}_{j'}(t')$. Each factor of ${\bf u}_{j'}$ in 
each term in the Taylor expansion is given by Eq. (\ref{C2}). We then make 
a Taylor series expansion of the ${\bf f(\bf R}_{j'}+{\bf u}_{j'}(t')-
{\bf v} t')$
 which occurs in each of these expressions. At any given point in this 
procedure, we can stop it by replacing 
${\bf f}({\bf R}_{j'}+{\bf u}_{j'}(t')-{\bf v} t')$ by its zeroth order 
term in 
the Taylor series expansion, namely ${\bf f}({\bf R}_{j'}-{\bf v} t')$. 
Then the 
resulting perturbation expansion of Eq. (\ref{C2}) can be represented 
schematically by the diagram in Fig. \ref{tree}. 
\noindent Each vertex is labeled by a position 
and time $({\bf R}_{j_n},t_n)$, where n is equal to 1 at the vertex 
that is 
furthest to the left. It increases as one moves to the right. 
 With the exception of the vertex 
at the beginning of the "tree" (i.e., the far left hand side of the 
diagram) each of these variables is summed and integrated over. A line 
connecting two vertices labeled by $({\bf R}_{j_1},t_1)$ and 
$({\bf R}_{j_2},t_2)$, for example, represents the Green's function 
${\bf G}({\bf R}_{j_1}-{\bf R}_{j_2},t_1-t_2)$, and each vertex denotes 
$(q-2)^{nd}$ order derivative of ${\bf f}$ multiplied by a factor of 
$1/(q-2)!$, where q is the number of lines intersecting at the vertex. 
It is 
not difficult to show that the last element in a chain of vertices 
(i.e., a 
line of vertices, each one of which has only two neighbors), which ends 
at the 
vertex $({\bf R}_{j_1},t_1)$ is a function of ${\bf R}_1$ and $t_1$ only 
in the 
combination $({\bf R}_{j_1}-{\bf v} t_1)$. For example, in d-dimensions 
this 
element is proportional to 
\begin{equation}
\label{C4}\int d^d k{e^{i{\bf k}\cdot ({\bf R}_1-{\bf v} t)}
{\bf f}({\bf k})\cdot{\bf \epsilon}_{{\bf k}}\over \omega^2 ({\bf k})
-({\bf v}\cdot{\bf k})^2+
 i\gamma{\bf v}\cdot{\bf k}}.
\end{equation}
In two and three dimensions the above integral can be done by contour 
integration. In the small v limit $\omega ({\bf k})$ can be approximated 
by $v_p k'$, 
where $v_p$ is the sound velocity and ${\bf k}={\bf G}+{\bf k'}$, where 
${\bf G}$ 
is the nearest reciprocal lattice vector to the region in which 
$\omega ({\bf k})$ and ${\bf v}\cdot{\bf k}$ intersect. Then the 
above integral becomes
\begin{equation}
\label{C5}\int d^d k' {{e^{i{\bf k}\cdot ({\bf R}_1-{\bf v} t)}
{\bf f} ({\bf G})\cdot{\bf \epsilon}_{{\bf G}}
\over v_p^2 k'^2-({\bf v}\cdot{\bf G})^2+i\gamma{\bf v}\cdot{\bf G}}}.
\end{equation}
In one dimension this integral is proportional to 
\begin{equation}
\label{C6}cos(vG(x_1-vt_1))e^{-\gamma |x_1-vt_1|}.
\end{equation}
In three dimensions, it is proportional to
\begin{equation}
\label {C6a} i{e^{i({\bf v}\cdot{\bf G}/v_p)|{\bf R}_1-{\bf v} t|}
e^{-(\gamma/v_p)|{\bf R}_1-{\bf v} t_1|}\over |{\bf R}_1-{\bf v} t_1|}.
\end{equation}
By building a chain out of these 
elements and integrating over each of the vertices except the furthest 
to the 
left, we find that if $({\bf R}_{j_1},t_1)$ are the labels of the 
end of the 
chain farthest to the left, the chain will be a function of these variable 
in the combination 
$({\bf R}_{j_1}-{\bf v} t_1)$. On doing the integrals and sums on each 
of the 
vertices in the chain except the last one, the Fourier transform of the 
$n^{th}$ Green's function in 
the resulting contribution to the perturbation theory is 
\begin{equation}
\label{C7} [\omega^2 ({\bf k}_n)-({\bf v}\cdot{\bf k}_n)^2+
i\gamma ({\bf v}\cdot{\bf k}_n)]^{-1}
\end{equation}
where ${\bf k}_n$ is one of the wavevectors which are summed over. Each 
vertex 
represents the dot product of one of the phonon polarization vectors 
$\hat{\bf \epsilon}_{{\bf k}_n}$ and the spatial Fourier transform of one of 
the functions f, evaluated at the difference between the wavevectors of the 
Green's functions on either side of the vertex, multiplied by the dot 
product of a polarization vector and the argument of the Fourier 
transforms of f appearing in the chain. Using the fact that a chain ending 
at the vertex labeled by $({\bf R}_{j_1},t_1)$ is a function of the variable 
$({\bf R}_{j_1}-{\bf v} t_1)$, it is easy to show that a vertex at which 
q chains intersect will also have this dependence on the space and time 
variables with which it is labeled, and it will also contain a factor 
\begin{equation}
\label{C8}\Pi_{m=1}^n [\omega^2 ({\bf k}_m)-({\bf v}\cdot{\bf k}_m)^2+
i\gamma {\bf k}_m)]^{-1}, 
\end{equation}
where n is the order in perturbation theory. Each variable ${\bf k}_m$ 
is summed 
over. When we take the thermodynamic limit for the film, each of these 
summations becomes a d-dimensional integral over the wavevector. 
If these integrals diverge at small v, the most divergent part will 
come from the integral over the vicinity of the 
intersection of the 
plane $\omega={\bf v}\cdot{\bf k}_m$ with the phonon dispersion surface 
$\omega({\bf k}_m)$. Therefore, let us consider the integral over just 
this region of k-space. Then 
in the small v limit, $\omega({\bf k}_m)$ can be 
expanded in a Taylor series about the film reciprocal lattice vector 
${\bf G}$ of the dip in $\omega({\bf k}_m)$ which is intersected by 
the plane $\omega={\bf v}\cdot{\bf k}_m$. Then, if we write ${\bf k}_m=
{\bf G}+{\bf k'}_m$, ${\bf v}\cdot{\bf k}_m$ and 
$\omega^2 ({\bf k}_m)$ are, to lowest order in ${\bf k}'_m$, ${\bf v}\cdot
{\bf G}$ 
and $(v_p k'_{m})^2$, respectively. Then the divergent part of one of 
these integrals is proportional to 
\begin{equation}
\label{C9} \int {d^d k' \over v_p^2 k'^2-({\bf v}\cdot{\bf G})^2+
i\gamma {\bf v}\cdot{\bf G}}.
\end{equation}
Let us consider this integral in the limit as $\gamma$ approaches zero 
since it is expected that the phonon damping will always be small 
compared to the frequency of the phonon being excited (which is equal to 
${\bf v}\cdot{\bf G}$). 
If we make a change of integration variable to a variable ${\bf x}$ equal 
to $v_p{\bf k}/(vG)$ the integral becomes 
\begin{equation}
\label{C10} (vG/v_p)^d (v_p/vG)^2 \int {d^d x 
e^{i{\bf x}\cdot ({\bf R}_j-{\bf v} t)}
\over x^2-1+i\epsilon},
\end{equation}
where $\epsilon=\gamma/vG$, which we are assuming to be $<<1$, since the 
damping of an acoustic phonon mode is always small compared to its frequency. 
The smallest 
value of x in the range of integration is 0 and the largest value is 
large compared to 1, but still finite. As long as $\epsilon$ is small 
compared to 1 but nonzero, the integral over x gives a finite number. Then 
the $n^{th}$ order term in the perturbation theory diverges as $v^{d-2}$ 
if $d<2$. Two dimensions is a marginal dimensionality, which could still 
have 
logarithmic divergences in v at small v.

\end{multicols}{2

\end{document}